\documentclass[a4paper]{jpconf}
\usepackage{graphicx}
\usepackage[T2A]{fontenc}
\usepackage[russian, english]{babel}
\usepackage{epstopdf}

\begin{document}
\title{High-energy atmospheric muon flux calculations in
comparison with recent measurements}

\author{A A Kochanov$^{1,2}$, A D Morozova$^{3,4}$,
	T S Sinegovskaya$^{5}$ and  S~I~Sinegovsky$^{2,4}$}
\address{$^1$ Institute of Solar-Terrestrial Physics SB RAS,  Lermontov str. 126а, Irkutsk, 664033, Russia}
\address{$^2$ Irkutsk State University, Gagarin blvd. 20, Irkutsk, 664003, Russia}
\address{$^3$ Lomonosov Moscow State University, Leninskie Gory 1,  Moscow, 119991, Russia}
\address{$^4$ Joint Institute for Nuclear Research, Joliot-Curie 6, Dubna, Moscow region, 141980, Russia}
\address{$^5$ Irkutsk State Transport University, Chernyshevskiy str. 15, Irkutsk, 664074, Russia}
%Dzhelepov Laboratory of Nuclear Problems, Joint Institute for Nuclear Research (DLNP, JINR)
%Faculty of Physics M.V. Lomonosov Moscow State University, Leninskie Gory,
%119991 Moscow, Russia

\ead{kochanov@iszf.irk.ru, refriz@yandex.ru, tanya@api.isu.ru, sinegovsky@jinr.ru}

\begin{abstract}
Recently the atmospheric muon spectra at high energies were reconstructed for two ranges of zenith angles,
basing on the events collected with the IceCube detector. These measurements reach high energies at which the contribution to atmospheric muon fluxes from decays of short-lived hadrons is expected.  
Latest IceCube measurements of the high-energy atmospheric muon spectrum indicate the presence of prompt muon component at energies above 500 TeV.

In this work, the atmospheric conventional muon flux in the energy range 10 GeV\,--\,10 PeV  is calculated using a set of hadronic models in combination with known parameterizations of the cosmic ray spectrum by Zatsepin \& Sokolskaya and by Hillas \& Gaisser.  The calculation of the prompt muons  with use of the quark-gluon string model (QGSM)  reproduces the muon data of IceCube experiment. % confirming the IceCube observations.
%The calculation of the prompt muons  with use of the quark-gluon string model (QGSM)  reproduces the %measured muon flux and confirming the IceCube observations.
Nevertheless,  an additional contribution to the prompt muon component is required to describe the IceCube muon spectra, if a charm production model predicts the appreciably lower prompt lepton flux as compared with QGSM.  This addition apparently originating from rare decay modes of the short-lived unflavored mesons $\eta, \eta^\prime, \rho, \omega, \phi$,  might ensure the competing contribution to the high-energy atmospheric muon flux. 
	
\end{abstract}

\section{Introduction}
High-energy atmospheric muons and neutrinos arise in weak decays of mesons produced in cosmic ray collisions with the Earth's atmosphere. Atmospheric muon flux is a tool to study the cosmic ray spectrum and elemental composition if there is a reliable model for hadron-nucleus interactions at high energies.  
Otherwise, if the elemental spectra of primary cosmic ray are well studied, a comparison of calculated  high-energy cosmic ray muon spectra with the experimental data allows one to get characteristics of the hadron-nucleus interactions. 

The atmospheric muon spectra at high energies were reconstructed for two ranges of zenith angles
basing on the events collected with IceCube detector \cite{IC_mu-aver}. These measurements are of strong interest because they reach the high energies at which atmospheric muons from decays of short-lived hadrons are expected.
The IceCube\,+\,IceTop combined analysis~\cite{IceTop17} resulted in the reconstruction of the high energy  muon spectrum separately, near vertical direction and close to horizontal, that  allows: (i) the comparison with the muon data of the ealier experiments; (ii) the search of a reliable model for high-energy hadronic interactions; (iii) the detection of the prompt muons component. 
 
We calculate the atmospheric muon spectrum in the energy range 10 GeV-10 PeV using the hadronic 
models  Kimel \& Mokhov \cite{km, kmn}, QGSJET-II \cite{qgsjet03, qgsjet04},  SIBYLL 2.1\cite{sib21}, EPOS-LHC \cite{epos15, epos17}, in combination with known parametrizations of 
the cosmic ray spectrum by Zatsepin \& Sokolskaya (ZS) \cite{ZS} and by Hillas \& Gaisser (H3a) \cite{H3a}. 
%Calculations agree with recent IceCube and IceTop measurements of the high-energy atmospheric muon flux.
The prompt muon contribution calculated with use of the quark-gluon string model (QGSM) \cite{bnsz} is compatible with recent IceCube measurements up to PeV energies. %range.
However in case of using the updated version of the QGSM \cite{ss17, ss18} or other  models of the charm production, which lead to appreciably lower prompt lepton flux, we find that an additional contribution to the prompt muon component is required to describe the IceCube muon spectra. 
This component can originate from rare decay modes of the short-lived unflavored vector mesons ($\eta, \eta^\prime, \rho, \omega, \phi$) \cite{volkova11, illana11} which contribute to the high-energy atmospheric muon flux.

\section{Methods of the calculation} 
The muon flux computations are performed with ${\cal Z}(E,h)$-method for solution of atmospheric hadron cascade equations \cite{ns00, kss08}. The method enables one to calculate atmospheric fluxes of mesons, nucleons, muons \cite{sks10, kss13} and neutrinos \cite{sms15} for non-power primary cosmic ray spectra, non-scaling inclusive cross sections and rising cross sections of inelastic hadron-nucleus collisions.
%for the non-power primary cosmic ray spectra, taking into account a violation of the Feynman scaling of inclusive cross sections and the rise with energy of inelastic cross sections for hadron-nucleus collisions.
Besides, we have performed the calculation with different approach, Matrix Cascade Equations (\cite{Fedyn15a, Fedyn15b}, using the free access package МCE{\scriptsize Q} \cite{MCEq17} (details of the comparison with ${\cal Z}(E,h)$-method see in \cite{mkss798, mkss934, Fed_sib23c}). 
%--------------------------Fig.1 -----------
The calculations are performed for a set of hadron-nucleus interaction models using the parametrizations of the primary spectrum based on experiments. Besides parametrizations of cosmic ray spectra by Zatsepin \& Sokolskaya and Hillas \& Gaisser (H3a) we use also the Nikolsky, Stamenov and Ushev  model (NSU) \cite{NSU} and the Elrykin, Krutikova and Shabelsky one (EKS) \cite{EKS}, as well as the toy model by Thunman, Ingelman and Gondolo (TIG) \cite{TIG}, in order to compare results of updated calculations with those of past years. 

\section{Energy spectra of atmospheric muons}
\begin{figure}[t!]
  \begin{center}
 \includegraphics[trim=0.5cm 0.5cm 0cm 1.0cm, width=0.60\textwidth] {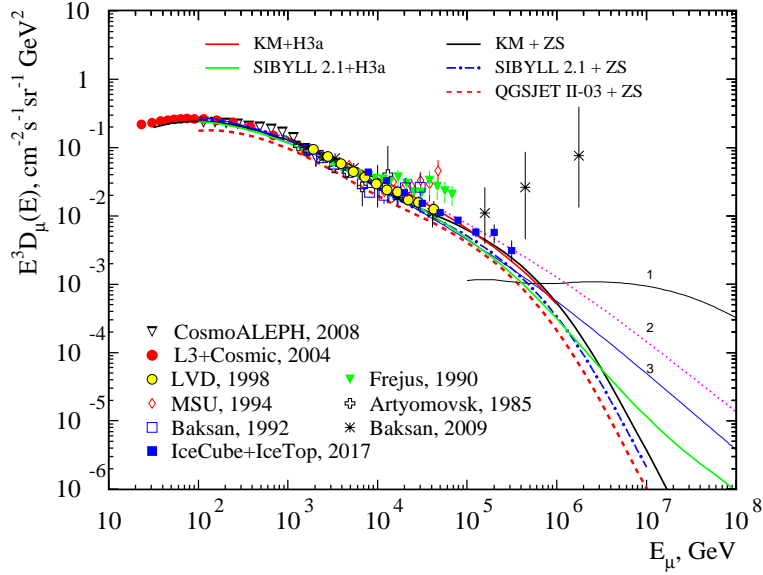}  %icetop_blue-4mod.eps
 % \vspace{-6pc}
  \caption{\label{IceTop_vert} Calculated muon flux at $\theta=0^\circ$ (curves) vs. the experimental  data of last decades measurements (see text) and latest results of the IceCube + IceTop recontruction \cite{IceTop17}.   
} 
   		\end{center}
	 \end{figure}

The  spectrum of atmospheric muons near the vertical is shown in figure \ref {IceTop_vert}. 
Data of last decades measurements at $\theta=0^\circ $  (points) and calculation results (set of curves) are presented. Experimental data are taken from Refs. \cite{kss08, sks10, kss13} (see references therein).
%Most expressive and ambiguous muon data (Baksan, 2009) were taken from Refs. \cite {bust09, bust12}. 
Muon data of the BUST experiment (Baksan, 2009) were taken from Refs. \cite {bust09, bust12}.
Earlier predictions for the conventional muons \cite{kss08} are represented by lines 2, 3 obtained  with the comsic ray spectra NSU and EKS correspondingly. Line 1 illustrates the prompt muons contribution calculated  with the early version of QGSM \cite{bnsz}.
 The rest lines show the calculations with hadronic models Kimel \& Mokhov (КМ), SIBYLL 2.1,  QGSJET II-03 combined with CR spectra by Zatsepin \& Sokolskaya (ZS) and by Hillas \& Gaisser (H3a). 
 Filled small squares represent the recent results of the combined IceCube\,+\,IceTop measurements~\cite{IceTop17}.
%--------------------------Fig.2 -----------
%--------------------------Fig.2 -----------
\begin{figure}[t]
  		\begin{center}
		    \includegraphics[trim=0.5cm 0.5cm 0cm 0cm, width=0.60\textwidth]{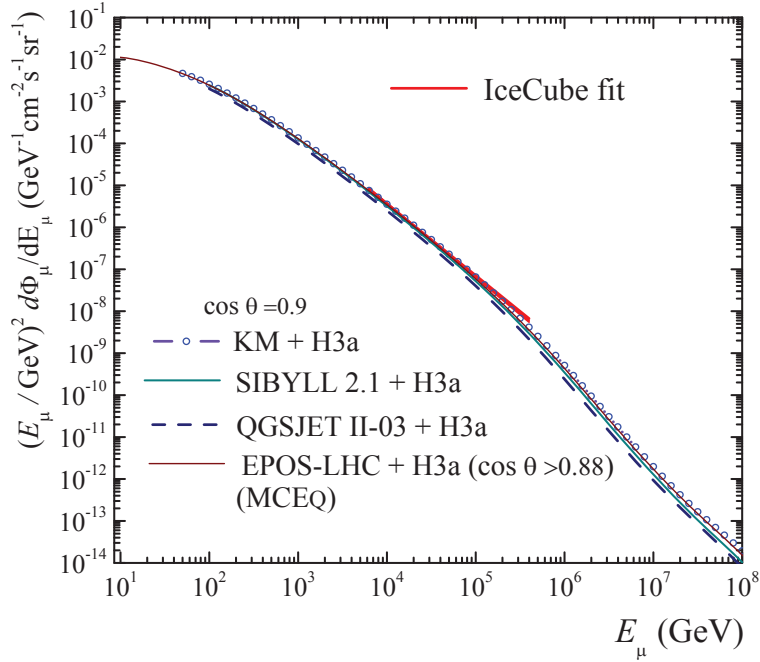}                                                                %{mu_EPOS_noPM.pdf}%{IceTop_vert.pdf} 
\caption{\label{IceTop_fit} Calculated muon spectrum near the vertical and the best fit \cite{IceTop17} for IceCube\,+\,IceTop data in the energy range $6-400$ TeV (red line).  Curves:  EPOS-LHC, КM, and SIBYLL 2.1 combined with the cosmic-ray spectrum H3a.}
	  		\end{center}
	 \end{figure} 
%--------------------------Fig.3 -----------
\begin{figure}[ht]
  		\begin{center}
		    \includegraphics[trim=0.5cm 0cm 0cm 0.5cm, width=0.60\textwidth] {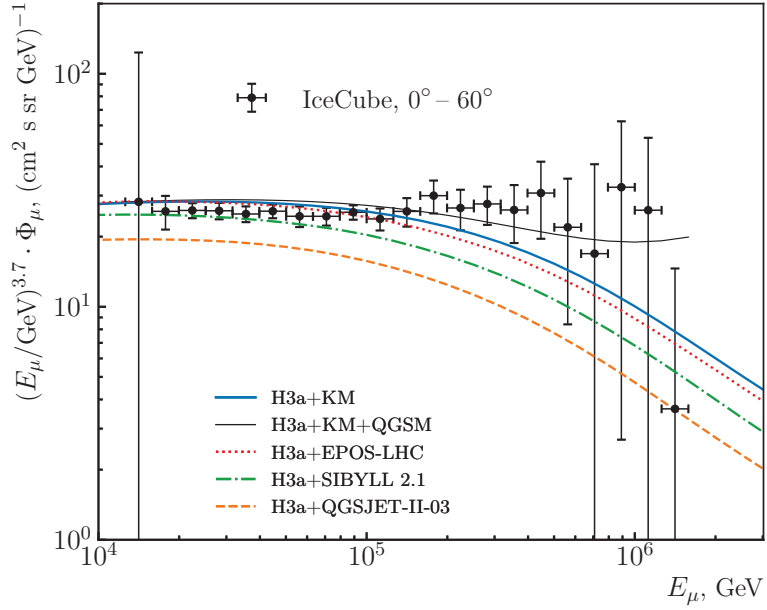}   
\caption{\label{fig3} IceCube data measurements \cite{IC_mu-aver} for zenith angles  $\theta < 60^\circ$. Curves: calculations for hadronic models  QGSJET II-03, SIBYLL 2.1, EPOS-LHC and КM combined with the cosmic-ray spectrum H3a.}   \end{center}
	 \end{figure}
%--------------------- Fig.4------------
%\par
\begin{figure}[t!]   %\vspace{0.5cm} 
  		\begin{center}
		    \includegraphics[trim=0.5cm 0cm 0cm 0cm, width=0.60\textwidth] {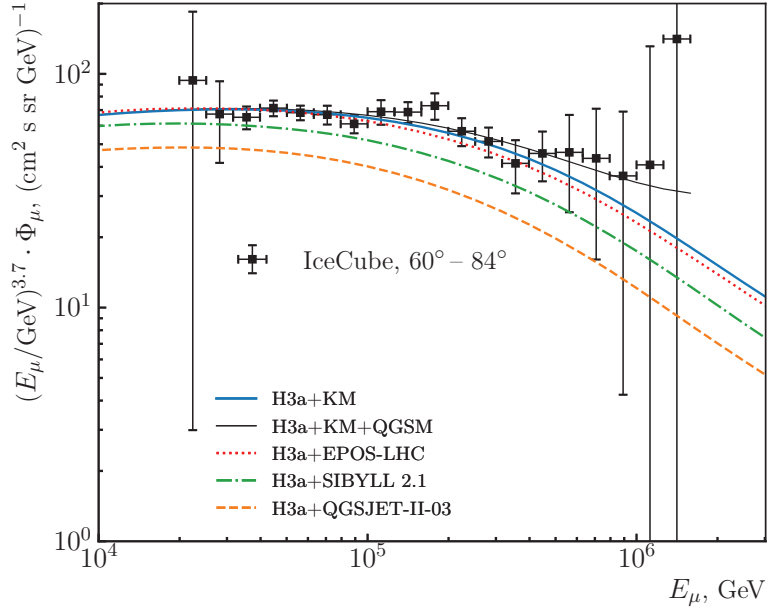}  % Fig4_hor.eps{IceCube_hor_v2.eps} %{IC_60-84v2.eps}  %{IC_60-84.pdf} % IceCube_60-84.pdf 
\caption{\label{IC_hor} IceCube data measurements for $\theta > 60^\circ$ and calculations.
The same notation as in figure \ref{fig3}.}
 	\end{center}                                                                     
	 \end{figure}      
	 %
%------------------ Fig.5 --------------		
	 \begin{figure}[ht]
  		\begin{center}          
		    \includegraphics[trim=0.5cm 0.5cm 0cm 1.0cm, width=0.60\textwidth] {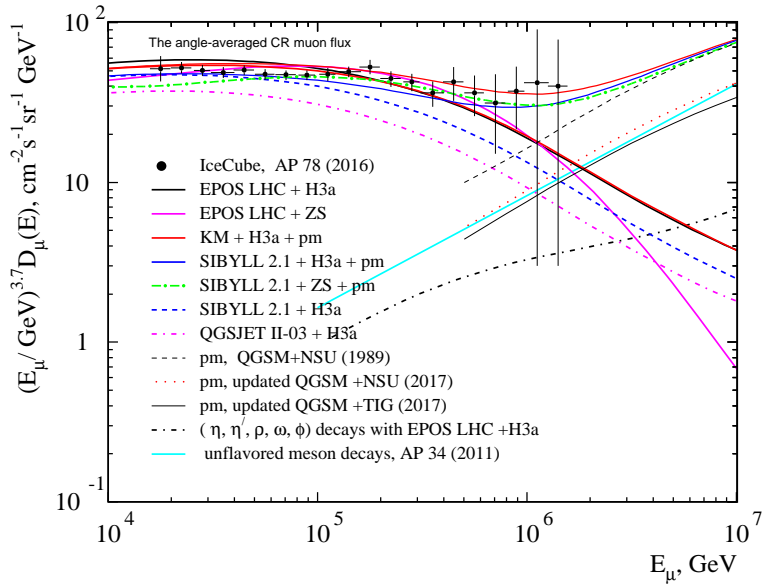} %{muratio.jpg}%%%  
  \caption{\label{all-sky} IceCube data (points) and calculations using EPOS-LHC, КM, SIBYLL 2.1, and QGSJET II-03 combined with cosmic-ray spectra models H3a and ZS.	Contribution of vector mesons decays ($\eta, \eta^\prime, \rho, \omega, \phi$): solid cyan line --  \cite{illana11}; black dash-dotted -- this work (EPOS-LHC).} 	
  		\end{center}
	 \end{figure}	
\par
%	\cite{IceTop17} $cos\theta > 0.88$ 
In figure \ref{IceTop_fit}, the calculated muon spectrum is compared with the best fit \cite{IceTop17} for IceCube\,+\,IceTop data, valid in the energy range $6-400$ TeV for near vertical directions ($\cos\theta > 0.88$) (solid line):  
\[\frac{d\Phi_\mu}{dE_\mu} = (9.0 \pm 0.3) \cdot 10^{-17} \left(\frac{E_\mu}
{50\ {\rm{TeV}}}\right)^{-3.74\pm 0.03}, \ {\rm{cm}^{-2} \rm{s}^{-1} \rm{sr}^{-1} \rm{GeV}^{-1}}.
\]		
Our calculations with EPOS-LHC and КM combined with H3a spectrum lead to the muon spectrum compatible with the fit, excepting the energy range beyond $200$ TeV.
\par 

The IceCube muon spectra \cite{IC_mu-aver} for zenith angles  $\theta < 60^\circ$ and  $\theta > 60^\circ$  are shown in figures \ref{fig3}, \ref{IC_hor} along with our calculations.
These figures  display the clear evidence in favor of the muon component with harder spectrum  than is expected for the conventional flux.  Such behavior is consistent with sizable contribution of prompt muons, and thus the IceCube experiment reveals the hard component, inspite of large errors.
As can be seen from the calculated curves in figure \ref{fig3}, at energies above $100$ TeV the slope of the conventional muon spectrum increases by $\sim 0.3$  due to the knee of the cosmic ray spectrum.  At angles $\theta > 60^\circ$ (figure \ref{IC_hor}), the difference between expected conventional flux and measured one is less as compared with that in figure \ref{fig3}, because of zenitn-angle enhancement of the conventional muon flux.  The prompt muon component calculated with QGSM, being added to the  conventional flux, agrees well with the IceCube data in both angle intervals. 
 	
The all-sky reconstruction of the atmospheric muon spectrum in IceCube experiment \cite{IC_mu-aver} is shown in figure \ref {all-sky}.  The set of curves shows the muon spectra calculated using models EPOS-LHC, КM, SIBYLL 2.1, and QGSJET II-03  with the CR spectra ZS and H3a. Also shown are the preceding QGSM calculations \cite{bnsz, sks10} of the prompt muon flux (black dashed line) and that with updated version of QGSM \cite{ss18}, performed for the same choice of the CR spectrum (NSU) (red dotted), and beside, for the TIG  spectrum (black thin line). The upper three curves present the sum of conventional and prompt muons (pm) \cite{sks10}: the red solid (top) line corresponds the model  KM\,+\,H3a\,+\,pm; blue solid -- the same for SIBYLL 2.1; green dash-dotted --  SIBYLL~2.1\,+\,ZS\,+\,pm. The result derived for EPOS-LHC (black solid line represents the conventional flux) is close to that for  КМ model. 
Solid cyan line corresponds to the contribution from decays of the unflavored mesons, taken from Ref. \cite{illana11} (SIBYLL 2.1\,+\,TIG). Black dash-dotted line corresponds to this work calculation with EPOS-LHC. 	

We may summarize: the calculation of the all-sky muon spectrum with use of models KM or EPOS-LHC (conventional flux) in combination with QGSM (prompt muons), describes  well the IceCube measurement data.

\section{Conclusions}
The comparison of the calculated high-energy atmospheric muon spectra with recent measurements
of IceCube experiment, as well as with prior ones, justifies the proper consideration of the muon production
in hadron showers, and reliability of the performed calculations. 
Atmosperic muon spectra calculated with hadronic models, EPOS-LHC, Kimel \& Mokhov, SIBYLL 2.1 are consistent with recent experimental results.  The computation for EPOS-LHC combined with the  Hillas \& Gaisser cosmic ray spectrum leads to closest agreement with the best fit of IceCube \cite{IceTop17} in the energy range $6\,-\,400$ TeV.                          
\par
High-energy spectra of atmospheric muons for two zenith-angle ranges, reconstructed in IceCube experiment,  evidence the prompt muons contribution. In spite of rather large errors, the experimental points against the background of the conventional atmospheric muons unambiguously indicate the presence of prompt muons at energies beyond 500 TeV. The discovery of the prompt  muon component  is the most important result of the latest  measurement of the cosmic-ray muon spectrum in IceCube experiment.

The prompt muon flux calculated with use of the quark-gluon string model \cite{bnsz} is compatible with these IceCube measurements up to PeV region.  However, usage of the updated version of the QGSM \cite{ss17,ss18}, as well as  different ``low-flux''  models of the charm production, leads to the appreciably lower prompt lepton flux as compared with Ref.~\cite{bnsz}. In that case, an additional contribution to the prompt muon component is required to describe the IceCube measured muon spectra. 

\section*{Acknowledgements}

This work was supported partly by the Russian Federation Ministry of  Science and Higher Education,
agreement 3.9678.2017/8.9.  A. Kochanov is supported with budgetary funding of Basic Research
program II.16 at ISTP SB RAS.


\section*{References}
\begin{thebibliography}{99}

\bibitem{IC_mu-aver}  
 Aartsen M G {\it et al} (IceCube Collaboration) 2016 {\it Astropart. Phys.} {\bf 78} 1-27 
%Characterization of the atmospheric muon flux in IceCube, p. 1-27  // 

\bibitem{IceTop17}
Tenholt F {\it et al} (IceCube Collaboration) 2017 High-energy atmospheric muons in IceCube and IceTop
35th Int. Cosmic Ray Conf.  (ICRC2017 Bexco, Busan, Korea) PoS (ICRC2017) 317 
({\it Preprint} 1710.01194)

\bibitem{km} Kimel L R and Mokhov N V 1974 {\it Izv. Vyssh. Uchebn. Zaved.} {\bf 10} 17
\bibitem{kmn} Kalinovsky A N, Mokhov N V and Nikitin Yu P 1989 {\it Passage of high-energy particles through matter} (New York: AIP)

\bibitem{qgsjet03} Ostapchenko S 2006 {\it Phys. Rev.} D {\bf 74} 014026
\bibitem{qgsjet04} Ostapchenko S 2011 {\it Phys. Rev.} D {\bf 83} 014018  
% %; Ostapchenko S 2008 {\it Nucl. Phys.} B (Proc. Suppl.) {\bf 175-176} 73 
\bibitem{sib21} Ahn E J, Engel R, Gaisser T K, Lipari P, Stanev T 2009 {\it Phys. Rev.} D {\bf 80} 094003

\bibitem{epos15} Pierog T,  Karpenko I,  Katzy J M, Yatsenko E  and  Werner K 2015 {\it Phys. Rev.} C {\bf 92} 034906 %({\it Preprint} 1306.0121)
\bibitem{epos17} Pierog T 2017 {\it EPJ Web Conf.}  {\bf 145} 18002 
      
\bibitem{ZS} Zatsepin V I and Sokolskaya N V 2006 {\it Astron. Astrophys.} {\bf 458} 1-5 
\bibitem{H3a} Gaisser T K  2012 {\it Astropart. Phys.}  {\bf 35} 801-806
%T.~Gaisser, Astropart. Phys.  24 (2012) 801-806; arXiv:1303.1431v1 [hep-ph].

\bibitem{bnsz} Bugaev E V, Naumov V A, Sinegovsky S I and Zaslavskaya E S 1989 {\it Nuovo Cim.} C {\bf 12} 41-73
\bibitem{ss17} Sinegovsky S I and  Sorokovikov M N 2017 {\it Rus. Phys. J.}  {\bf 60} 1189 
\bibitem{ss18} Sinegovsky S I and Sorokovikov M N 2018 Prompt atmospheric neutrinos in the quark-gluon string model {\it Preprint} P2-2018-4  JINR, Dubna

\bibitem{volkova11}
Volkova L V 2011 {\it Phys. Atom. Nucl.} {\bf 74} 318-323
%Cosmic-ray muons at ultrahigh energies
% ЯФ. 2011. Т.74. C. 336–341.
%МЮОНЫ КОСМИЧЕСКИХ ЛУЧЕЙ ПРИ СВЕРХВЫСОКИХ ЭНЕРГИЯХ
%

\bibitem{illana11}
Illana J I, Lipari P, Masip M and Meloni D 2011 {\it Astropart. Phys.} {\bf 34} 663-673
%Atmospheric muon and neutrino fluxes at very high energy~/ J. I. Illana et. al. 


\bibitem{ns00}  Naumov V A and Sinegovskaya T S 2000 {\it Phys. Atom. Nucl.} {\bf 63}  1927-1935
\bibitem{kss08} Kochanov A A, Sinegovskaya T S and Sinegovsky S I 2008 {\it  Astropart. Phys.} {\bf 30} 219
\bibitem{sks10} Sinegovsky S I, Kochanov A A, T.S. Sinegovskaya, Misaki A  and  Takahashi N  2010 {\it Int. J. Mod. Phys.} A {\bf 25}  3733  %3733-3740 

\bibitem{kss13} Kochanov A A, Sinegovskaya T S and Sinegovsky S I 2013  {\it J. Exp. Theor. Phys.}  {\bf 116} 395  
\bibitem{sms15}	Sinegovskaya T S,  Morozova A D and  Sinegovsky S I 2015  {\it Phys. Rev.} D {\bf 91} 063011

\bibitem{Fedyn15a}  Fedynitch A,  Engel R, Gaisser T K, Riehn F and  Stanev T 2015 Calculation of conventional and prompt lepton fluxes at very high energy {\it EPJ Web Conf.} {\bf 99} 08001 ({\it Preprint} 1503.00544)  
\bibitem{Fedyn15b}  Fedynitch A,  Engel R, Gaisser T K, Riehn F and  Stanev T 2015  {\it PoS} (ICRC 2015)  1129  % MCEq
\bibitem{MCEq17}   Fedynitch A 2017  https://github.com/afedynitch/MCEq 

\bibitem{mkss798}  Morozova A D, Kochanov A A, Sinegovskaya T S and Sinegovsky S I  2017 {\it J. Phys. Conf. Ser.} {\bf 798}, 012101
\bibitem{mkss934}  Morozova A D, Kochanov A A, Sinegovskaya T S and Sinegovsky S I 2017  {\it J. Phys. Conf. Ser.} {\bf 934}, 012008

\bibitem{Fed_sib23c} Fedynitch A,  Engel R, Gaisser T K, Riehn F and Stanev T 2018 
The hadronic interaction model Sibyll 2.3c and inclusive lepton fluxes {\it Preprint} 1806.04140

\bibitem{NSU}
Nikolsky S I, Stamenov J N and Ushev S Z 1984 {\it Sov. Phys. JETP} {\bf 60} 10-21
\bibitem{EKS}
Erlykin A D, Krutikova N P and Shabelskii Yu M 1987 {\it Yad. Fiz.} {\bf 45} 1075 
\bibitem{TIG}
Thunman M, Ingelman G and Gondolo P 1996 {\it Astropart. Phys.} {\bf 5} 309

\bibitem{bust09}
Bogdanov A G, Kokoulin R P, Novoseltsev Yu F, Novoseltseva R V, Petkov V B and Petrukhin A A 2009
Energy spectrum of cosmic ray muons in 100 TeV energy region reconstructed from the BUST data
{\it Preprint 0911.1692} %v1 [astro-ph.HE],  9 Nov 2009.

\bibitem{bust12} Bogdanov A G, Kokoulin R P, Novoseltsev Yu F, Novoseltseva R V, Petkov V B and Petrukhin A A 2012 {\it Astropart. Phys.}  {\bf 36} 224
%Energy spectrum of cosmic ray muons in 100 TeV energy region reconstructed from the BUST data

\end{thebibliography}
\end{document}